\documentclass[%
 reprint,
amsmath,amssymb,
aps,
prl,
twocolumn,
]{revtex4-2}
\usepackage{braket}
\usepackage{graphicx}
\usepackage{bm}
\usepackage{float}
\usepackage{color}
\usepackage{hyperref}
\usepackage{placeins}

\usepackage[normalem]{ulem}

\begin{document}

\title{Non-Pauli errors can be efficiently sampled in qudit surface codes}
\author{Yue Ma$^1$}

\author{Michael Hanks$^1$}

\author{M. S. Kim$^1$}

\affiliation{$^1$QOLS, Blackett Laboratory, Imperial College London, London SW7 2AZ, United Kingdom\\
}

\begin{abstract}

Surface codes are the most promising candidates for fault-tolerant quantum computation. Single qudit errors are typically modelled as Pauli operators, to which general errors are converted via randomizing methods. In this Letter, we quantify remaining correlations after syndrome measurement for a qudit 2D surface code subject to non-Pauli errors. Using belief propagation and percolation theory, we relate correlations to loops on the lattice. Below the error correction threshold, remaining correlations are sparse and locally constrained. Syndromes for qudit surface codes are therefore efficiently samplable for non-Pauli errors, independent of the exact forms of the error and decoder.

\end{abstract}

\maketitle

Quantum error correction is an important element towards large-scale quantum computation~\cite{shor1995scheme,calderbank1996good,steane1996multiple,raussendorf2012key,campbell2017roads,girvin2021introduction}. Topological codes on a two-dimensional plane~\cite{bravyi1998quantum,dennis2002topological,kitaev2003fault,freedman2001projective,wang2011surface,fowler2012surface,stephens2014fault,terhal2015quantum} such as the toric code and the surface code, are among the most widely studied codes, as they only involve local interaction of the quantum registers~\cite{terhal2015quantum}. A variety of experiments have demonstrated the path towards building a qubit topological code~\cite{barends2014superconducting,nigg2014quantum,kelly2015state,corcoles2015demonstration,takita2016demonstration,erhard2021entangling,chen2021exponential,marques2022logical,krinner2022realizing,zhao2022realization}.
Recent studies have shown computational advantages using $d$-dimensional qudit systems~\cite{bullock2005asymptotically,bocharov2017factoring,pavlidis2021quantum,gustafson2022noise}.
The 2D toric code and surface code have been generalized to qudits~\cite{bullock2007qudit},
and although experimental manipulations are more challenging for these systems, significant progress~\cite{lanyon2008manipulating,bianchetti2010control,wang2018multidimensional,ringbauer2018certification,blok2021quantum,morvan2021qutrit,yurtalan2020implementation,ringbauer2022universal,chi2022programmable,hrmo2022native,goss2022high,roy2022realization}
in qudit control has been made, and it has also been numerically shown that a larger qudit dimension can lead to an increased, more tolerant error correction threshold~\cite{duclos2013kitaev,anwar2014fast,andrist2015error,watson2015fast,hutter2015improved,marks2017comparison}.

Most studies of stabilizer error correction codes focus on incoherent Pauli errors~\cite{scruby2022non}, considering their classical simulability~\cite{gottesman1998heisenberg} and justified by the discretization of errors by stabilizer measurements~\cite{shor1995scheme,nielsen_chuang_2010,raussendorf2012key}.
However, increasing technological precision has motivated the study of more realistic error models in qubit codes~\cite{ghosh2012surface,darmawan2017tensor,bravyi2018correcting,beale2018quantum,huang2019performance,iverson2020coherence,venn2020error,bonilla2021xzzx,scruby2022non,tiurev2022correcting,novais2013surface,brown2016quantum}. Broader classes of channels that can be classically efficiently simulated have been investigated~\cite{magesan2013modeling,gutierrez2013approximation,puzzuoli2014tractable}, and the Pauli Twirling Approximation (PTA) turns out to be a practical approach to mapping the non-Pauli error channel to the code-compatible Pauli model~\cite{silva2008scalable,geller2013efficient,tomita2014low,gutierrez2015comparison,katabarwa2015logical,katabarwa2017dynamical,gutierrez2016errors,cai2019constructing,martinez2020approximating}.
Despite the effort in qubit codes, general error models for qudit codes have received less attention~\cite{grassl2018quantum}, and the implementation of qudit Pauli twirling~\cite{jafarzadeh2020randomized,goswami2021quantum} requires a larger set of twirling gates as the dimension $d$ of the qudit increases, making it operationally difficult. Moreover, it has been shown that in both transmon qutrit~\cite{goss2022high,morvan2021qutrit,blok2021quantum} and trapped ion qudit~\cite{klimov2003qutrit,low2020practical} systems the naturally occurring errors are not the cyclic shifts described by the most common generalization of the Pauli matrices. These considerations call into question the benefit of explicitly applying the Pauli twirling gates, and lead us to ask, to what extent can non-Pauli error models be investigated, such that their intrinsic correlations may be leveraged to improve the codes?

In this Letter, we study the qudit 2D square lattice surface code, where each physical qudit has the same probability of having experienced an error described by a general unitary operator. By sampling the syndromes
and classifying the error subgraphs~\cite{Diestel2017,Bondy2010,xu2014geometric}, inspired by the notion of belief propagation in computer science~\cite{pearl2022reverend,kim1983computational}, we connect the problem of the remaining correlations removed in the PTA with the distribution of loop sizes in Bernoulli percolation theory~\cite{bazant2000largest}.
We find that for a physical error rate below the qudit error correction threshold, with an upper limit estimated at $30\%$~\cite{anwar2014fast}, the partially discretized coherent errors are sparse and mostly involve only four neighboring qudits, while the largest spatial span of their correlation is local, growing only logarithmically with the code distance. As the computational complexity depends exponentially on the span of these correlations, the strong localization observed implies that explicitly keeping track of correlations remains numerically efficient.
Our findings, independent of dimension and the specific forms of the non-Pauli errors, justify the possibility of efficiently exploring qudit code performances beyond Pauli error models.

\begin{figure}[t]
\centering
\includegraphics[width=0.48\textwidth]{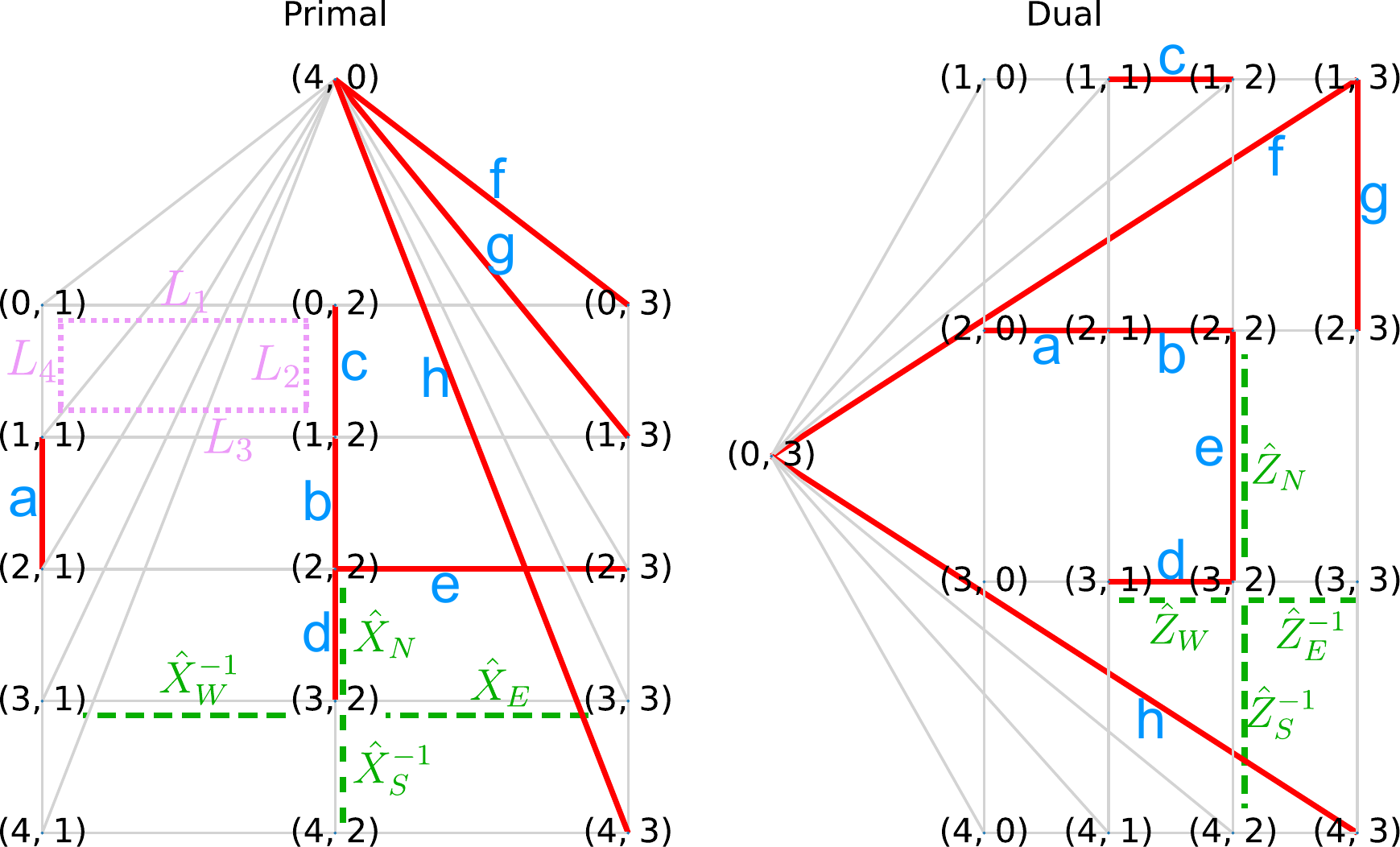}
\caption{An example 2D surface code with ${n_h=4}$ and ${n_v=5}$.
On the primal lattice, the horizontal (vertical) size $n_h$ ($n_v$) refers to the number of edges (nodes) in each row (column). Qudits subject to the error $\hat{F}$ are marked in thick red lines and labelled by blue letters. An example of a vertex [plaquette] operator is shown on the primal [dual] lattice centered at the node (3,2) [(3,2)], where relevant qudits on the edges are marked by green dashed lines and the Pauli operators are labelled in green. An example loop is shown in pink dotted lines, with the qudits labelled as $L_1\rightarrow L_4$. }\label{fig:lattice}
\end{figure}

\noindent \textit{Surface code error discretization} --- Consider the two-dimensional surface code with $d$-dimensional qudits on a square lattice~\cite{watson2015fast}, with size characterized by the two parameters $n_h$ and $n_v$. An example is shown in Fig.~\ref{fig:lattice}. On the primal lattice, the vertical (horizontal) boundaries are smooth (rough). For later convenience when dealing with deformed stabilizers at the boundaries and logical operators, we contract all the nodes on rough boundaries into a single dummy-node.
The $d$-dimensional Pauli operators are ${\hat{X}=\sum_{j=0}^{d-1}|j\oplus1\rangle\langle j|}$ and ${\hat{Z}=\sum_{j=0}^{d-1}\omega^j|j\rangle\langle j|}$, where $\oplus$ is the sum modulo $d$ and ${\omega=\exp(i2\pi/d)}$. They satisfy the commutation relation $\hat{X}^j\hat{Z}^k=\omega^{-jk}\hat{Z}^k\hat{X}^j$.
Stabilizer generators are vertex operators in the Pauli-$X$ basis and plaquette operators in the Pauli-$Z$ basis, though for our purposes the $Z$-basis stabilisers are better viewed as vertex operators on the dual lattice.

We assume a quantum channel $\mathcal{L}_s$ for each qudit,
\begin{equation}
    \mathcal{L}_s(\rho)=(1-p)\rho+p\hat{F}\rho\hat{F}^{\dagger},\ \mathrm{with}\ \hat{F}=\sum_{i,j=0}^{d-1}f_{i,j}\hat{X}^i\hat{Z}^j.
\end{equation}
Each qudit experiences a general unitary error $\hat{F}$ with probability $p$.
This operator can be decomposed into a linear sum of Pauli operators, as $\{\hat{X}^i\hat{Z}^j\}$ forms the complete Heisenberg-Weyl basis~\cite{asadian2016heisenberg}.
An example error pattern is shown in Fig.~\ref{fig:lattice}, where the erroneous qudit edges are marked in red and labelled $a\rightarrow h$.
These edges induce the \emph{error subgraph}.

For each error pattern, syndrome measurements project the erroneous state into a subspace compatible with a specific syndrome~\cite{raussendorf2012key}. As the syndrome subspace is constructed via Pauli operators acting on the code space, we expand the product of the non-Pauli $\hat{F}$ errors into a sum of multi-qudit Pauli operators.
Importantly, if two multi-qudit Pauli operators bring the logical state into the same syndrome subspace, the probability of being projected into this subspace will depend on interference between the complex amplitudes $f_{i,j}$.
This effect, absent under the PTA, increases the computational complexity of numerically sampling the syndrome.
Intuitively, this effect appears if the error pattern can lead to the formation of a stabilizer or a logical operator.
On the contracted lattice (e.g. Fig.~\ref{fig:lattice}), this corresponds to the formation of loops,
which we discuss below.

We first argue that for each error pattern, if the error subgraphs on both the contracted primal and dual lattices are forests, the syndrome measurements fully discretize the errors in the Pauli basis. A \textit{forest} is a graph that contains no loops, and each connected component in a forest is a \textit{tree} \cite{Diestel2017,Bondy2010}.
Our argument is based on observation of the single-error case, followed by induction using the generalized Bayes' rule for belief propagation in a tree graph~\cite{pearl2022reverend,kim1983computational}.
Consider the primal lattice in Fig.~\ref{fig:lattice}.
The error subgraph contains three trees: $\{\rm{a}\}$, $\{\rm{b,c,d,e}\}$ and $\{\rm{f,g,h}\}$.
For each tree, we begin with stabilizer measurements at leaf nodes such as the node $(0,2)$.
Measurement discretizes the error operator for the qudit on the adjacent edge (qudit $\rm{c}$).
We then continue with sequential measurements along the branches towards the root node $(2,2)$.
The tree structure guarantees that measurements centered at leaf nodes are independent of each other, and that at most one adjacent edge corresponds to an error that has not yet been fully discretized by previous measurements~\cite{pearl2022reverend}.
Where trees contain the contracted node (such as the tree $\{\rm{f,g,h}\}$), this node is always the root and there is no associated measurement.

Complete discretization of errors in the Pauli basis suggests that the error channel can be replaced by the Pauli twirled channel,
\begin{equation}\label{eq:channelTwirl}    \tilde{\mathcal{L}}_s(\rho)=\frac{1}{d^2}\sum_{i,j=0}^{d-1}\hat{Z}^{-j}\hat{X}^{-i}\mathcal{L}_s(\hat{X}^i\hat{Z}^j\rho\hat{Z}^{-j}\hat{X}^{-i})\hat{X}^i\hat{Z}^j,
\end{equation}
whose effect is to keep only the diagonal terms in $\hat{F}\rho\hat{F}^{\dagger}$.
Importantly, this equivalence only relies on the relative positions of the erroneous qudits, and is independent of the qudit dimension $d$ and the specific values of $f_{i,j}$ in $\hat{F}$.
It is straightforward to generalize this equivalence to an arbitrary mixture of single-qudit errors applying some $\hat{F}^{(k)}$ with probability $p_k$, and even allowing different errors for each qudit.
From an operational perspective, this equivalence implies that for forest error subgraphs the computational complexity is the same as that of the PTA channel Eq.~\eqref{eq:channelTwirl}.

The error subgraph may contain loops. While edges that do not form part of a loop can be fully discretized following the above argument, errors on loop-edges can only be partially discretized in the Pauli basis (see the Supplemental Material~\cite{supp} for an explicit example). 
Consider a loop, for example, consisting of qudits $L_1\rightarrow L_4$ marked in pink dashed lines in Fig.~\ref{fig:lattice}.
The error $\hat{F}_{L_1}\hat{F}_{L_2}\hat{F}_{L_3}\hat{F}_{L_4}$ can be expanded as a sum over products of Pauli operators, which contains the stabilizer operator $\hat{S}_{p}=\hat{Z}_{L_1}\hat{Z}_{L_2}^{-1}\hat{Z}_{L_3}^{-1}\hat{Z}_{L_4}$ and all of its powers $\hat{S}_{p}^{k}$ for $k=0,1,\cdots,d-1$.
Products of Pauli errors that differ only by a power of $\hat{S}_{p}$ correspond to the same syndrome:
The probability of projection into the compatible subspace differs from the PTA due to the interference of complex amplitudes $f_{i,j}$ on the loop.
For each connected component in the error subgraph, the numerical cost is expected to increase with $d^{l}$, where $l$ is the tree-width of the component \cite{robertson_graph_1986,jensen_bayesian_1990,markov2008simulating}.
We can therefore study the increase in sampling complexity by quantifying the number and size of loops in the error subgraph, independent of the specific form of the error $\hat{F}$.
Once the syndrome is sampled, Pauli operators compatible with the syndrome are applied to recover the logical state and decoders developed for Pauli error models can be applied.

\noindent \textit{Relation to percolation theory} --- The 2D surface code modelled above can be directly mapped into the problem of Bernoulli percolation~\cite{duminil2018sixty} on the square lattice, which is a bond percolation model~\cite{stauffer2018introduction} where each edge is retained only with probability $p$, independent of other edges.
Bernoulli percolation has been extensively studied, and size distributions for clusters (connected components of the error subgraph) have been investigated both above and below the critical probability ${p_{c}=0.5}$~\cite{bazant2000largest,grinchuk2002large}.
In the sub-critical regime, for instance, the mean size of the largest connected cluster has been found to scale logarithmically with the total number of edges.
However, to the best of our knowledge only at the percolation threshold ~\cite{stanley1977cluster,herrmann1984building,gyure1995mass,bunde2000percolation,he2021size} has the distribution of loops been studied~\cite{herrmann1984building,xu2014geometric}.
We are interested in this distribution below the error correction threshold~\cite{anwar2014fast}, which is in the sub-critical percolation regime.
We use two measures to infer the loop size distribution.
The first is the proportion of error subgraph edges that are loop-edges, which quantifies how likely loops are to form (scaling with $p$ also providing information on the mean size). The second measure is the maximal one-dimensional span of a loop. This quantifies the size of the largest loop, in contrast to the mean, that can form in the sub-critical lattice.

\begin{figure}[t]
\centering
\includegraphics[width=0.48\textwidth]{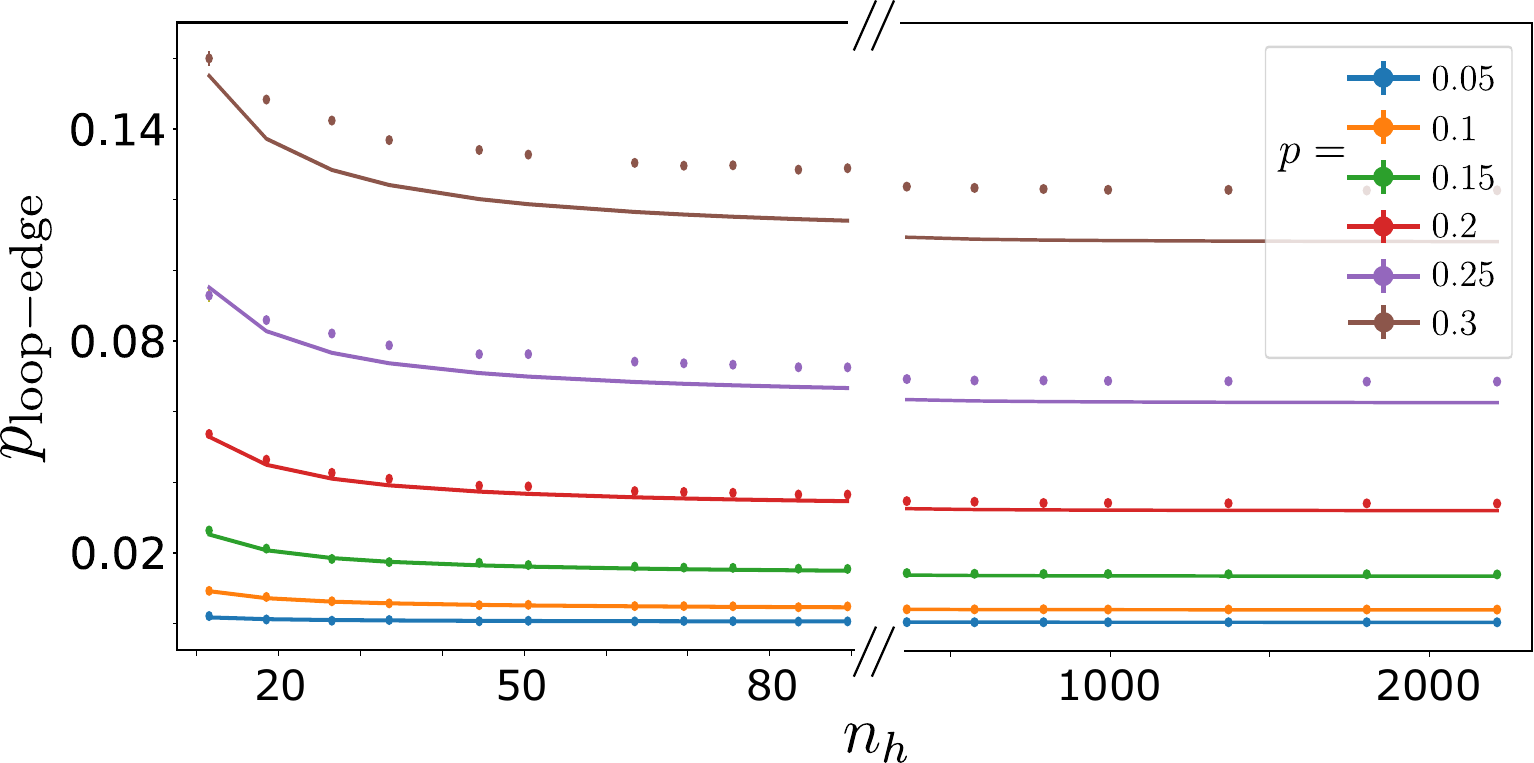}
\caption{The proportion of erroneous qudits in loops as a function of the lattice size $n_h$, for near-symmetric lattices $n_h=n_v\pm1$.
Colors correspond to different single qudit physical error rates. Solid lines follow the analytical expression Eq.~\eqref{eq:ratio_ana}. Markers are from the Monte-Carlo simulations, and  error bars represent the standard deviation. The left (right) half of the figure shows the behavior for smaller (larger) values of $n_h$, for a Monte-Carlo sample size of $2000$ ($200$).}\label{fig:Ratio_sym}
\end{figure}

\begin{figure}[t]
\centering
\includegraphics[width=0.48\textwidth]{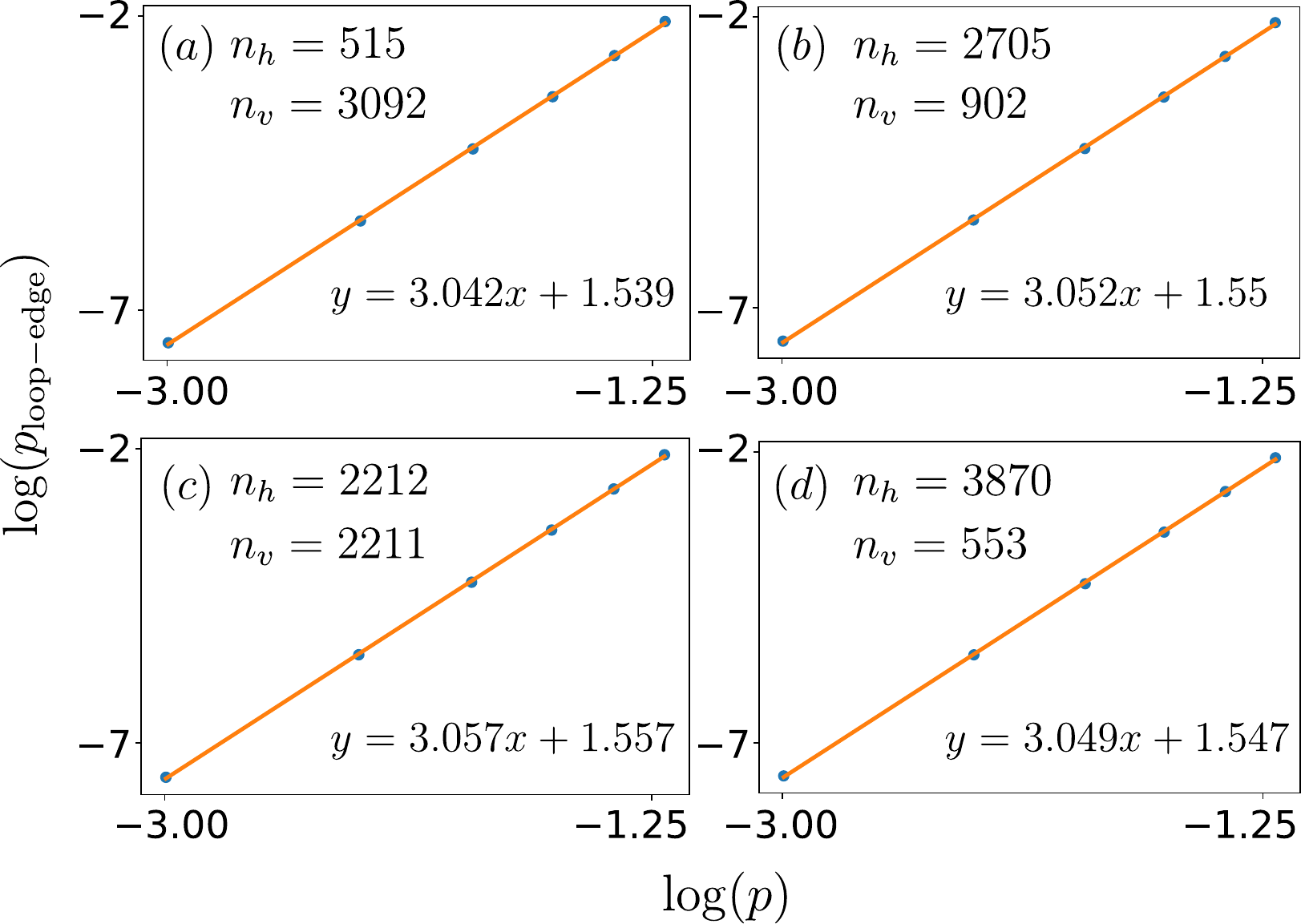}
\caption{Linear fitting of $y=\log(p_{\mathrm{loop-edge}})$ as a function of $x=\log(p)$ for four different lattice sizes. Blue markers are for the numerical data from the Monte-Carlo simulations, while orange lines are the results of the fitting (expressions shown).
}\label{fig:Ratio_fit}
\end{figure}

\noindent \textit{Proportion of loop-edges} --- We first consider the ratio between the number of qudits associated with loop-edges on the error subgraphs and the total number of erroneous qudits, denoted $p_{\mathrm{loop-edge}}$.
We use Monte-Carlo simulations, finding a cycle-basis in the error subgraph~\cite{SciPyProceedings_11,paton1969algorithm} for each sampled error pattern on the primal lattice.
This process is repeated for the dual lattice, removing redundancies where a qudit is involved in loops on both lattices and taking the sample average.

We can obtain an approximate analytical formula for $p_{\mathrm{loop-edge}}$. Below the percolation threshold, the dominant contributions come from the smallest loops, which include the three-edge loops on the rough boundaries and the four-edge loops inside the bulk of each lattice.
The primal [or dual] lattice has $2(n_v-1)$ [or $2(n_h-1)$] three-edge loops, each appearing with probability roughly $p^3$, and $(n_h-2)(n_v-1)$ [or $(n_h-1)(n_v-2)$] four-edge loops, each appearing with probability roughly $p^4$.
We therefore get
\begin{equation}\label{eq:ratio_ana}
    p_{\mathrm{loop-edge}}\approx 3\frac{n_h+n_v}{n_hn_v}p^2+4p^3.
\end{equation}
The first part of Eq.~\eqref{eq:ratio_ana} describes the dependence of $p_{\mathrm{loop-edge}}$ on the size and shape of the surface code, which matters more for smaller lattices. The second part only depends on $p$, capturing the asymptotic behavior for larger lattices. In Fig.~\ref{fig:Ratio_sym}, we show the value of $p_{\mathrm{loop-edge}}$ as a function of $n_h$, for the most symmetric shapes, $n_h=n_v\pm1$.

The analytical expression Eq.~\eqref{eq:ratio_ana} fits well with the numerical simulation result for $p\leq0.2$, but for larger $p$ the analytical formula underestimates $p_{\mathrm{loop-edge}}$.
This implies that the contribution from larger loops becomes non-negligible as $p$ increases.
We consider a correction term in place of the $4p^3$ term in Eq.~\eqref{eq:ratio_ana}. As plotted in Fig.~\ref{fig:Ratio_fit}, for the four pairs of $(n_h,n_v)$ in the asymptotic regime, the linear functions of $\log(p_{\mathrm{loop-edge}})$ versus $\log(p)$ are very similar. This verifies the approximate shape-independence of $p_{\mathrm{loop-edge}}$, and indicates that we may take the simulation result of the surface code with the maximum number of physical qudits, which is panel (c) in Fig.~\ref{fig:Ratio_fit}, to define the correction as $4.745p^{3.057}$ that replaces $4p^3$ in Eq.~\eqref{eq:ratio_ana}.
The exponent $3.057$ implies that for $p>0.2$, though the average loop size is slightly larger than $4$, most of the correlations remain tightly constrained to the smallest local loops. 

We have also numerically verified that the difference between the analytical and numerical results of $p_{\mathrm{loop-edge}}$ can be compensated if instead of the linear fitting, we add a term $6p^5$ to the analytical expression Eq.~\eqref{eq:ratio_ana}.
This implies that in addition to the dominant four-edged loops, a non-negligible number of six-edged loops appear in the error subgraph.
Connected components with loops of at most four and six edges both have tree width $l=2$.
Based on this modified expression for $p_{\mathrm{loop-edge}}$, we can deduce that they appear approximately $2n_hn_vp^4$ and $2n_hn_vp^6$ times respectively for each error pattern.
As a result, the increased time complexity due to residual correlations is expected to be $O(2n_hn_v(p^4+p^6)d^2)$. This is polynomial in both the code distance and the qudit dimension with power $2$, indicating that numerical sampling remains efficient for lattices of the sizes considered, with $n_{h}$ into the thousands.

In the Supplemental Material~\cite{supp}, we demonstrate the dependence of $p_{\mathrm{loop-edge}}$ on the aspect ratio $n_h/n_v$. We also describe a further binary criterion based on the simple presence or absence of loops, which is especially useful for bounding the accuracy of logical error rate estimates for small lattices.

\begin{figure}[t]
\centering
\includegraphics[width=0.48\textwidth]{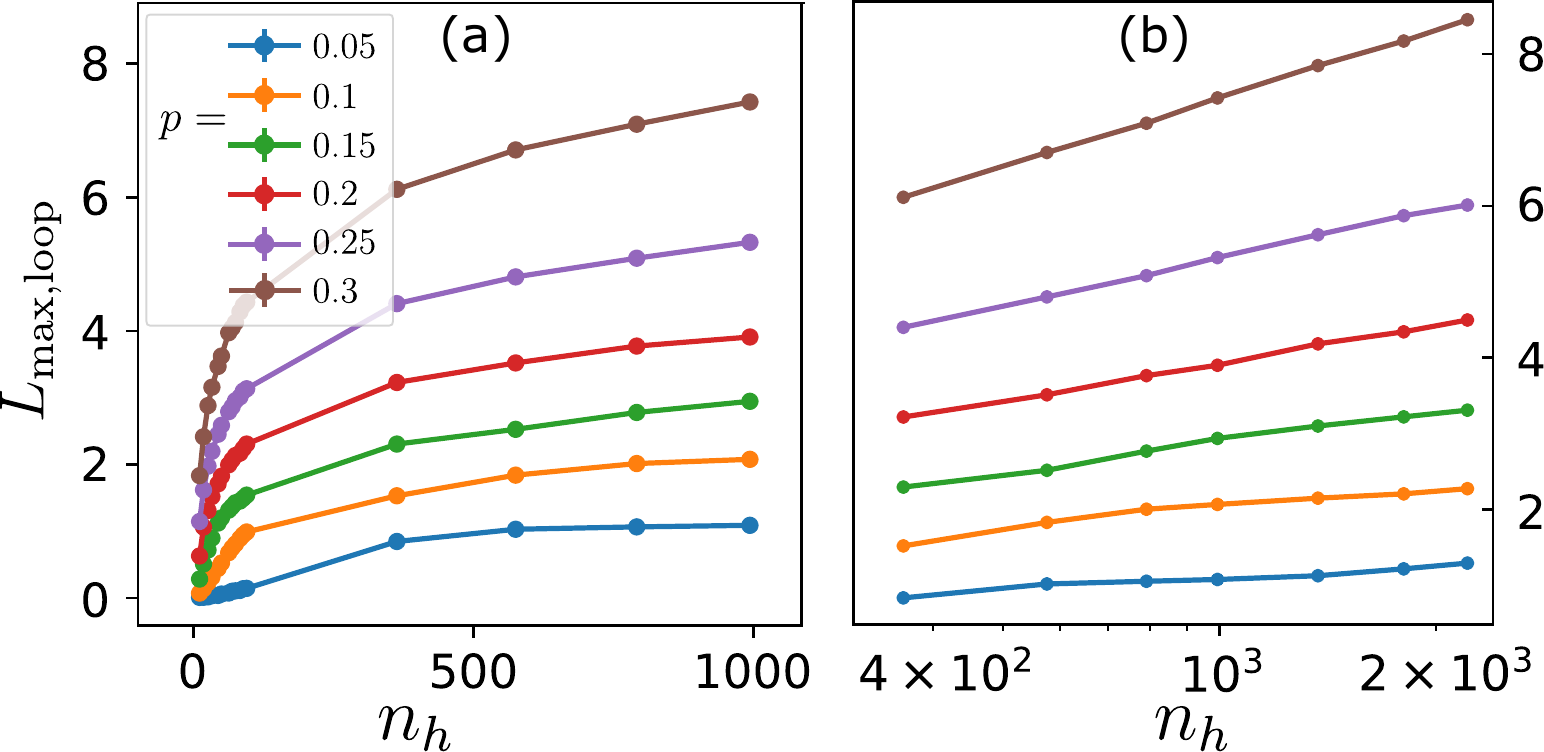}
\caption{The maximal one-dimensional span of a loop as a function of the lattice size $n_h$ for almost symmetric lattices $n_h=n_v\pm1$.
Points are obtained via Monte-Carlo simulation with 2000 samples, and error bars represent the standard deviation.
Colors correspond to different single qudit physical error rates. (a) Linear scale. (b) Logarithmic scale.}\label{fig:maxLoop}
\end{figure}

\noindent \textit{Maximal one-dimensional span of a loop} --- The proportion of loop-edges considered above describes the average behavior. Taking a different perspective, we next estimate the maximal loop size, putting an upper bound on the span of the correlations.
Specifically, for each loop we take the one-dimensional span, this being the maximal length of its smallest bounding rectangle. For example, for the four-edged elementary loop considered above, the span is 1. Taking the maximum span over all loops on both the primal and the dual lattices results in our target value, $L_{\mathrm{max,loop}}$. 

In Fig.~\ref{fig:maxLoop}, we show the numerical result of how $L_{\mathrm{max,loop}}$ behaves as a function of $n_h$ for symmetric lattices, on linear and logarithmic scales.
As $n_h$ increases, $L_{\mathrm{max,loop}}$ also increases but at a much slower rate.
The straight lines, especially for larger values of $p$, indicate that the maximal one-dimensional span of a loop grows logarithmically with the lattice size.
This scaling is similar to the finding in percolation theory that below the percolation threshold, the maximal cluster size increases logarithmically~\cite{bazant2000largest}, though the full cluster also includes non-loop edges.
Our finding focusing on loops only thus supplements the scaling behaviors in percolation theory.

As ${L_{\mathrm{max,loop}}+1}$ bounds the maximal tree width of a cluster, its logarithmic scaling with respect to $n_h$ indicates that the sampling complexity scales at most polynomially in the lattice size. As shown in Fig.~\ref{fig:maxLoop}, up to $n_h=2\times10^3$ and $p=0.3$, this complexity is on the order of $d^{9}$.
While the exponent increases very slowly with $n_h$, we do observe a rapid increase with $p$.
It was shown in Ref.~\cite{anwar2014fast} that $p=0.3$ corresponds to the accuracy threshold in the limit $d\rightarrow\infty$, while for more practical values of $d$ the threshold will be closer to $p=0.2$.
Therefore, for numerical investigations of qudit surface code thresholds we expect an upper bound on the sampling complexity closer to $d^{5}$. 

In the Supplemental Material~\cite{supp} we use another measure for the maximum loop size, the total number of nodes in a loop, with similar behavior to Fig.~\ref{fig:maxLoop}(a).
We also show that due to the slow increase of $L_{\mathrm{max,loop}}$ with $n_h$, lattice asymmetry has little effect.

\noindent \textit{Conclusion} ---
In this Letter, identifying that residual multi-qudit correlations in topological codes are restricted to cell boundaries (or loops, in 2D) of the error subgraph, we assess the computational complexity of syndrome sampling in qudit surface codes under stochastic single-qudit unitary error models from the perspectives of belief-propagation (or tensor-network contraction) and percolation theory.
Our analysis is independent of the qudit dimension and the specific form of the errors.

Considering a single-qudit physical error rate $p$ up to $30\%$ (the highest threshold predicted in qudit surface codes~\cite{anwar2014fast}), we quantify the average and maximal error loop sizes.
We find that the former, which as a mean value is related to the run time of the code simulation, is well modelled accounting only for four-edged loops plus a small correction from six-edged loops on the order of $p^2$, both of which have tree-width $2$.
The latter, which as an upper bound is related to the maximal memory required, grows only logarithmically with the code size $n_h$, dropping rapidly as $p$ decreases, and up to $n_h=2000$ and $p=0.3$ gives an upper bound of $9$ for the tree-width.

The efficient scaling observed in our results indicates that syndrome sampling complexity does not prevent the inclusion of non-Pauli errors; the complexity of the decoding step is expected to remain the limiting factor.
An investigation of qudit surface codes under general stochastic unitary errors is therefore within reach.
Intrinsic correlations may be leveraged to improve the codes, especially for $d>2$ qudit systems that have a richer structure than the biased error models of qubit systems.

By taking into account other percolation structures, our method may also be generalized beyond a single-qudit error model, and even to multi-qudit coherent errors~\cite{bravyi2018correcting}, where the discretization of errors by stabilizer measurements appears also as a central step.

\noindent \textit{Acknowledgements} ---
We acknowledge financial support from the Samsung GRC project, the UK Hub in Quantum Computing and Simulation
with funding from UKRI EPSRC grant EP/T001062/1, EPSRC Distributed Quantum Computing and Applications grant EP/W032643/1, and the UK National Quantum Hub for Imaging (QUANTIC, No. EP/T00097X/1).

\end{document}


\title{Supplemental Material: Non-Pauli errors can be efficiently sampled in qudit surface codes}
\author{Yue Ma$^1$}

\author{Michael Hanks$^1$}

\author{M. S. Kim$^1$}

\affiliation{$^1$QOLS, Blackett Laboratory, Imperial College London, London SW7 2AZ, United Kingdom\\
}





\maketitle

\section{Explicit examples of discretizing the errors by stabilizer measurements}

\begin{figure}[h]
\centering
\includegraphics[width=0.5\textwidth]{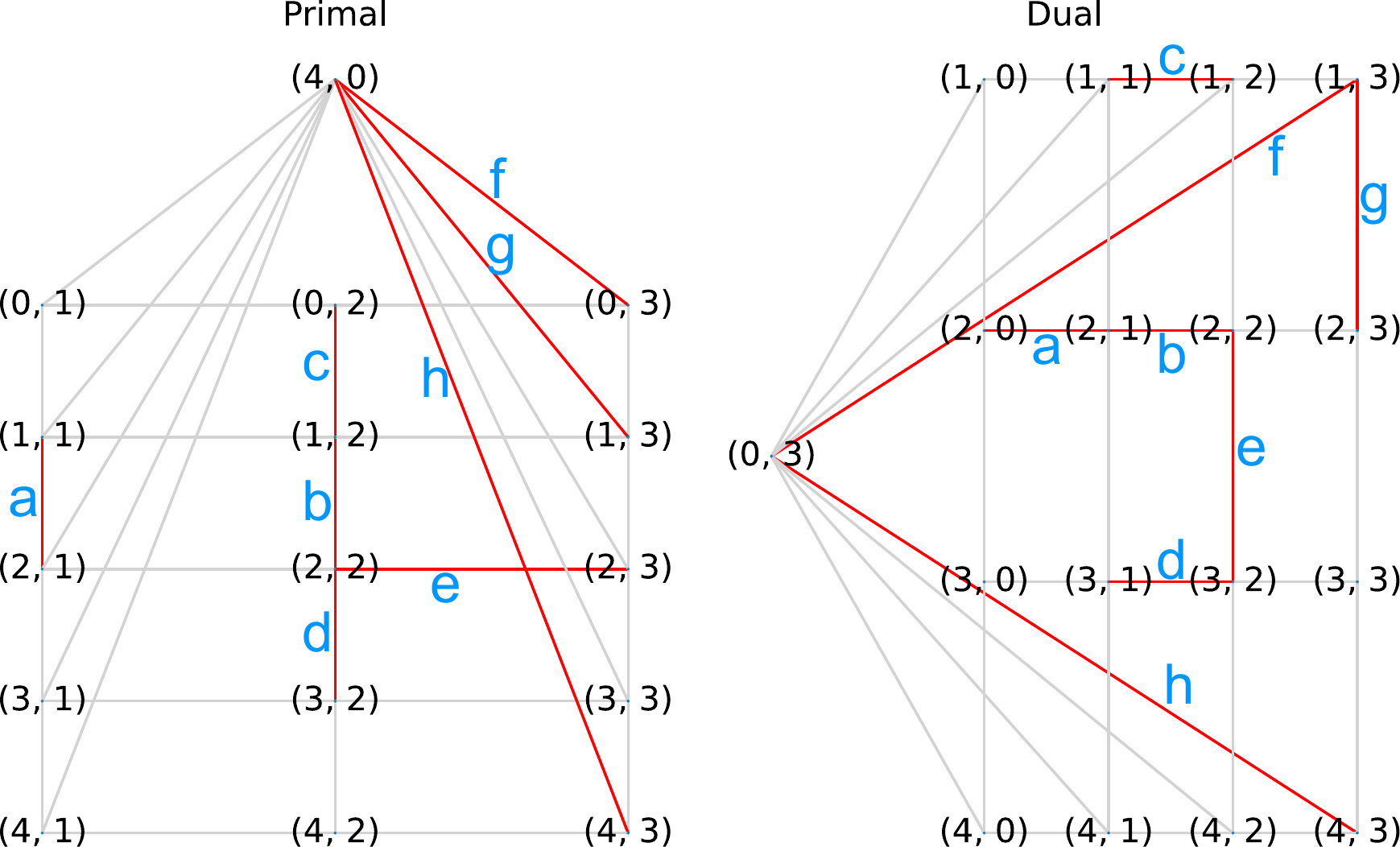}
\caption{An example of a 2D surface code with $n_h=4$ and $n_v=5$, after contracting the nodes on the rough boundaries to one node. Both the primal lattice and the dual lattice are shown. On the primal lattice, the horizontal size $n_h$ refers to the number of edges in each row, while the vertical size $n_v$ refers to the number of nodes in each column. Qudits that have been subject to the error $\hat{F}$ are marked in red and labelled by blue letters.}\label{fig:supp_lattice}
\end{figure}

As shown in Fig.~\ref{fig:supp_lattice}, we consider the error pattern where the edges that correspond to the qudits with error $\hat{F}$ are colored red and labelled a to h. On the primal lattice, the error subgraph consists of three disconnected components, none of which containing loops. The initial state is $\hat{F}_a\hat{F}_b\hat{F}_c\hat{F}_d\hat{F}_e\hat{F}_f\hat{F}_g\hat{F}_h|\psi\rangle$, where subscripts a to h label the qudits, and the state $|\psi\rangle$ is in the code space. We start from the isolated qudit with error, qudit a. We measure the vertex operator centered at node $(1,1)$, obtaining a measurement result $x_a$, which can take a value $0,1,\cdots$ or $d-1$, and correspondingly project the state to $\sum_{i=0}^{d-1}f_{i,x_a}\hat{X}_a^i\hat{Z}_a^{x_a}\hat{F}_b\hat{F}_c\hat{F}_d\hat{F}_e\hat{F}_f\hat{F}_g\hat{F}_h|\psi\rangle$. Note that the Pauli-$Z$ operator for qudit a now only has exponential $x_a$. We then measure the vertex operator centered at node $(2,1)$ to get the measurement result $-x_a$ and no change in the state. This additional stabilizer measurement with definite result and without changing the state indicates that the measurements are well-defined at the rough boundaries, which will be shown later. Next we focus on the tree of qudits with errors, labelled as b, c, d, e. The general procedure is to start measuring the vertex operator centered at one end node, and proceed along the chain until for the next vertex measurement there are more than one edge that have not been projected yet.
We stop there and start from another end node. Finally we can resume the measurement of vertices that were left before as more projections of edges have been implemented. The order of vertex operator measurements is: $(0,2)\rightarrow(1,2)\rightarrow(3,2)\rightarrow(2,3)\rightarrow(2,2)$. The measurement results are denoted as $x_c$, $x_b$, $x_d$, $x_e$, and $-x_c-x_b-x_e-x_d$, respectively, where $x_b,x_c,x_d,x_e$ can take a value from $0,1,\cdots,d-1$. The state is projected to $\sum_{i=0}^{d-1}f_{i,x_a}\hat{X}_a^i$ $\sum_{i=0}^{d-1}f_{i,x_c+x_b}\hat{X}_b^i$ $\sum_{i=0}^{d-1}f_{i,x_c}\hat{X}_c^i$ $\sum_{i=0}^{d-1}f_{i,-x_d}\hat{X}_d^i$ $\sum_{i=0}^{d-1}f_{i,x_e}\hat{X}_e^i$ $\hat{Z}_a^{x_a}$ $\hat{Z}_b^{x_c+x_b}$ $\hat{Z}_c^{x_c}$ $\hat{Z}_d^{-x_d}$ $\hat{Z}_e^{x_e}$ $\hat{F}_f$ $\hat{F}_g$ $\hat{F}_h$ $|\psi\rangle$ after the final measurement. To deal with the boundary qudits f, g and h which are only covered by one vertex measurement each, we can invert the contraction process such that they are described by $((0,3),(0,4))$, $((1,3),(1,4))$, $((4,3),(4,4))$, respectively. Measuring the vertex operators $(0,3)$, $(1,3)$ and $(4,3)$ correspondingly discretize the qudits f, g and h, and the state is projected to $\sum_{i=0}^{d-1}f_{i,x_a}\hat{X}_a^i$ $\sum_{i=0}^{d-1}f_{i,x_c+x_b}\hat{X}_b^i$ $\sum_{i=0}^{d-1}f_{i,x_c}\hat{X}_c^i$ $\sum_{i=0}^{d-1}f_{i,-x_d}\hat{X}_d^i$ $\sum_{i=0}^{d-1}f_{i,x_e}\hat{X}_e^i$ $\sum_{i=0}^{d-1}f_{i,-x_f}\hat{X}_f^i$ $\sum_{i=0}^{d-1}f_{i,-x_g}\hat{X}_g^i$ $\sum_{i=0}^{d-1}f_{i,-x_h}\hat{X}_h^i$ $\hat{Z}_a^{x_a}$ $\hat{Z}_b^{x_c+x_b}$ $\hat{Z}_c^{x_c}$ $\hat{Z}_d^{-x_d}$ $\hat{Z}_e^{x_e}$ $\hat{Z}_f^{-x_f}$ $\hat{Z}_g^{-x_g}$ 
 $\hat{Z}_h^{-x_h}$ $|\psi\rangle$, where all the $Z$ errors are fully discretized. Similarly, measurements of the plaquette operators discretize $X$ errors. As the error subgraph on the dual lattice in Fig.~\ref{fig:supp_lattice} also does not contain loops, the $X$ errors are fully discretized as well. We have therefore demonstrated the complete error discretization by syndrome measurements for a forest error subgraph,
\begin{align}           &\hat{P}^{(\cdots,x_a,\cdots,x_h,z_a,\cdots,z_h,\cdots)}\hat{F}_a\cdots\hat{F}_h|\psi\rangle\nonumber\\
&=f_{-z_a,x_a}f_{-z_a-z_b,x_c+x_b}f_{-z_c,x_c}f_{z_a+z_b+z_d+z_e,-x_d}\nonumber\\
&~~~~f_{-z_a-z_b-z_e,x_e}f_{z_f+z_g,-x_f}f_{z_g,-x_g}f_{-z_h,-x_h}\nonumber\\
&~~~~\hat{X}_a^{-z_a}\hat{Z}_a^{x_a}\hat{X}_b^{-z_a-z_b}\hat{Z}_b^{x_c+x_b}\hat{X}_c^{-z_c}\hat{Z}_c^{x_c}\hat{X}_d^{z_a+z_b+z_d+z_e}\hat{Z}_d^{-x_d}\nonumber\\
&~~~~\hat{X}_e^{-z_a-z_b-z_e}\hat{Z}_e^{x_e}\hat{X}_f^{z_f+z_g}\hat{Z}_f^{-x_f}\hat{X}_g^{z_g}\hat{Z}_g^{-x_g}\hat{X}_h^{-z_h}\hat{Z}_h^{-x_h}|\psi\rangle,
\end{align}
where $\hat{P}^{(\cdots)}$ is the projector corresponding to a specific set of syndromes. 

For each forest error subgraph, different Pauli errors correspond to distinct syndrome measurement outcomes. Upon relabelling the superscripts we can arrive at the equivalence with the Pauli twirled error channel, \begin{equation}\label{eq:supp_channelTwirl}
\mathcal{F}(\rho)=\sum_{i,j=0}^{d-1}|f_{i,j}|^2\hat{X}^i\hat{Z}^j\rho\hat{Z}^{-j}\hat{X}^{-i},
\end{equation}
as
\begin{align}
    &\hat{P}^{(\Gamma_{a,b,\cdots})}\cdots\hat{F}_b\hat{F}_a|\psi\rangle\langle\psi|\hat{F}_a^{\dagger}\hat{F}_b^{\dagger}\cdots\hat{P}^{(\Gamma_{a,b,\cdots})}=\nonumber\\
    &\hat{P}^{(\Gamma_{a,b,\cdots})}\cdots\mathcal{F}_b(\mathcal{F}_a(|\psi\rangle\langle\psi|))\cdots\hat{P}^{(\Gamma_{a,b,\cdots})}
\end{align}
for a forest error subgraph and $\Gamma_{a,b,\cdots}$ represents any compatible syndrome measurement outcome. 

\begin{figure}[t]
\centering
\includegraphics[width=0.5\textwidth]{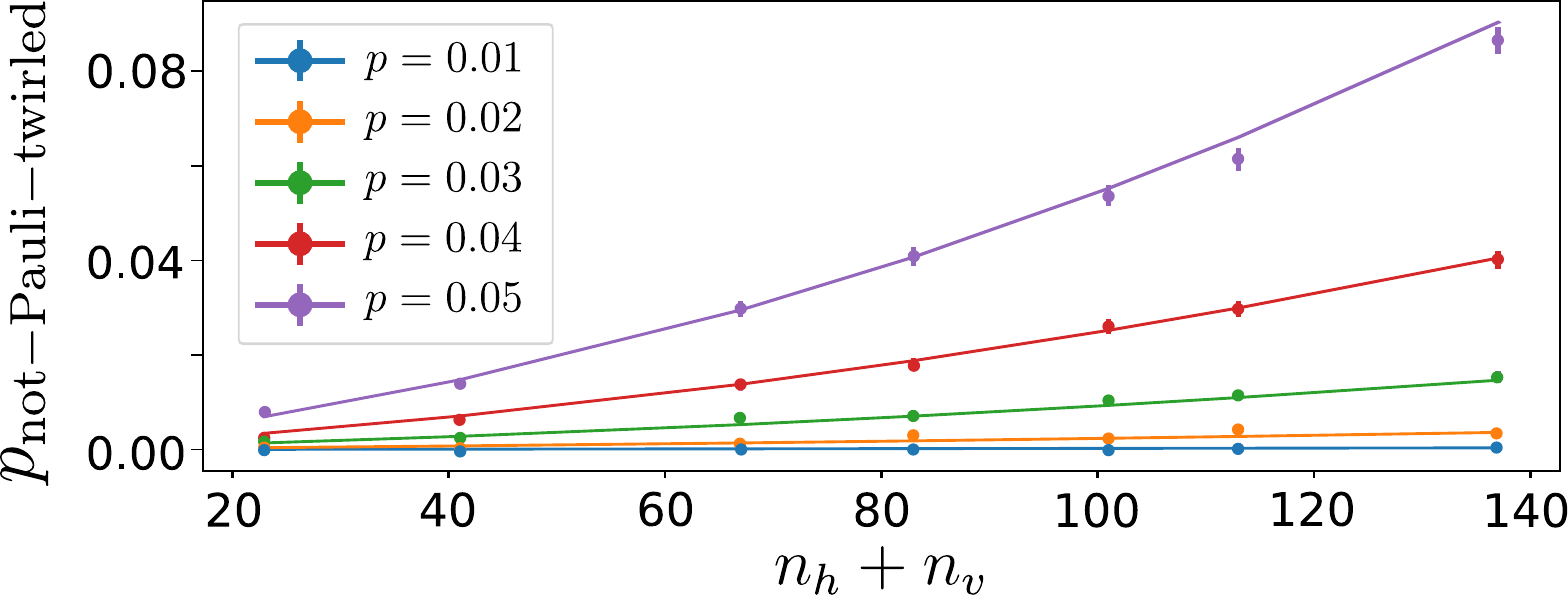}
\caption{The probability of not equivalent to the Pauli-twirled error model as a function of the lattice size characterized by $n_h+n_v$, for almost symmetric lattices $n_h=n_v\pm1$. Different colors correspond to different single qudit physical error rates. Solid lines follow the analytical expression Eq.~\eqref{eq:count_ana}. Markers are from the Monte-Carlo simulations, where for each data point $10000$ samples are taken. Error bars represent the standard deviation.}\label{fig:count_sym}
\end{figure}

The equivalence shown above does not hold if the error subgraph contains loops. As an example, we consider an error subgraph on the primal lattice formed by the nodes $(3,1)$, $(3,2)$, $(2,2)$ and $(2,1)$. We label the corresponding qudits with error as $A: ((3,1),(3,2))$, $B: ((3,2),(2,2))$, $C: ((2,2),(2,1))$ and $D: ((2,1),(3,1))$. Suppose the Pauli-$X$ errors have been completely discretized by the measurements of plaquette operators and the state is 
\begin{align}
&\hat{X}_A^{z_A}\hat{X}_B^{z_B}\hat{X}_C^{z_C}\hat{X}_D^{z_D}\sum_{j_A,j_B,j_C,j_D=0}^{d-1}f_{z_A,j_A}f_{z_B,j_B}f_{z_C,j_C}f_{z_D,j_D}\nonumber\\
&\hat{Z}_A^{j_A}\hat{Z}_B^{j_B}\hat{Z}_C^{j_C}\hat{Z}_D^{j_D}|\psi\rangle.\nonumber
\end{align}
The vertex operators are measured in the order of $(3,1)\rightarrow(3,2)\rightarrow(2,2)\rightarrow(2,1)$, and the outcomes are $x_A,x_B,x_C,-x_A-x_B-x_C$, respectively. The state is projected to
\begin{align}
    &\hat{X}_A^{z_A}\hat{X}_B^{z_B}\hat{X}_C^{z_C}\hat{X}_D^{z_D}\hat{Z}_B^{-x_B}\hat{Z}_C^{x_C+x_B}\hat{Z}_D^{-x_A}\nonumber\\
    &\sum_{j_A=0}^{d-1}f_{z_A,j_A}f_{z_B,-x_B+j_A}f_{z_C,x_C+x_B-j_A}f_{z_D,-x_A-j_A}|\psi\rangle.\nonumber
\end{align}
The remaining sum of the products of $f_{ij}$ coefficients is due to the fact that there is one less stabilizer measurement for a loop compared with a tree subgraph. It can also be explained as coming from the superposition of equivalent paths that are different by the application of a stabilizer operator, therefore corresponding to the same syndrome. It indicates that due to the correlations this situation cannot be represented by the Pauli-twirled channel Eq.~\eqref{eq:supp_channelTwirl}. We have contracted all the nodes on the rough boundaries to one node, as illustrated in Fig.~\ref{fig:supp_lattice}, to cover the situations of logical operators and boundary stabilizers as loops. For instance, on the primal lattice, the loop made of nodes $(4,0),(1,1),(1,2),(1,3)$ is a logical operator, and the loop made of nodes $(4,0),(0,1),(1,1)$ corresponds to a deformed stabilizer generator.

\section{Probability of not equivalent to the Pauli-twirled channel}

\begin{figure}[t]
\centering
\includegraphics[width=0.5\textwidth]{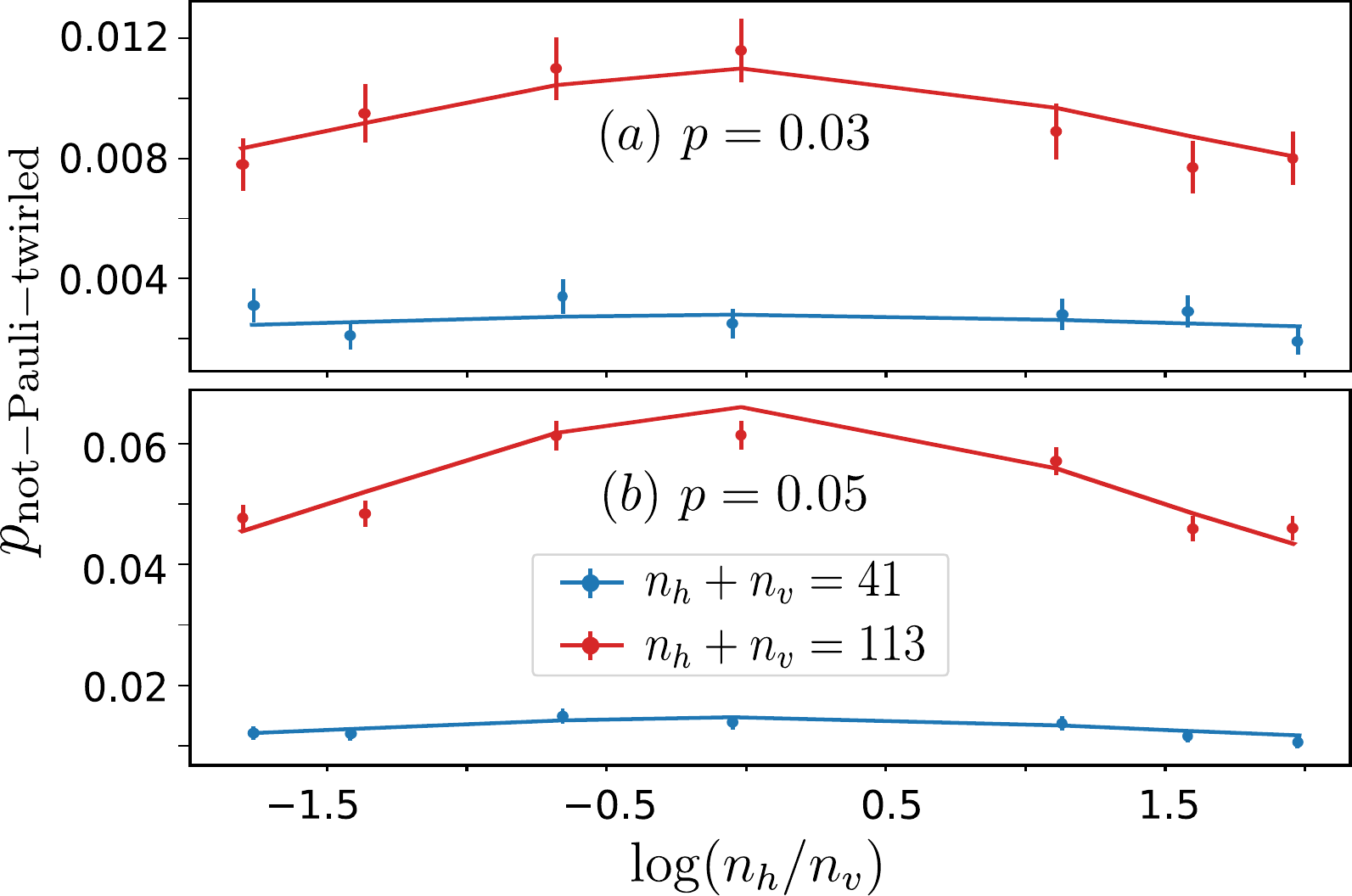}
\caption{The probability of not being equivalent to the Pauli-twirled error model as a function of the aspect ratio of the lattice, fixing $n_h+n_v=41$ (blue) or $n_h+n_v=113$ (red). We fix the single qudit physical error rate as (a) $p=0.03$, (b) $p=0.05$. Solid lines follow the analytical expression Eq.~\eqref{eq:count_ana}. Markers are from the Monte-Carlo simulations, and  error bars represent the standard deviation.}\label{fig:count_ana}
\end{figure}

We can estimate the probability of having error patterns that do not contain loops in both the primal lattice and the dual lattice. To achieve this, we assume each qudit is subject to the original error channel $\mathcal{L}_s$ [see main text Eq.~(1)] and sample the error patterns via the Monte-Carlo simulation. The probability that the result is not equivalent to the Pauli-twirled channel is estimated by the number of sampled error patterns that contains loops divided by the total number of samples.

We can also derive an approximate analytical expression to estimate the probability of having loops. As discussed in the main text, by taking into account only the smallest loops (three-edge on the rough boundaries and four-edge in the bulk), we have
\begin{equation}\label{eq:count_ana}
    p_{\mathrm{not-Pauli-twirled}}\approx n_hn_v\cdot 2p^4+(n_h+n_v)(2p^3-3p^4).
\end{equation}

In Fig.~\ref{fig:count_sym}, we illustrate $p_{\mathrm{not-Pauli-twirled}}$ as a function of the lattice size $n_h+n_v$, for almost symmetric lattice shapes $n_h=n_v\pm1$. The analytical formula Eq.~\eqref{eq:count_ana} fits well with the numerical results. For a fixed physical error rate $p$, a larger lattice corresponds to a larger probability of deviation from the Pauli-twirled error channel. This is because a larger number of qudits also indicate a larger number of error edges, therefore forming a loop in one of the error subgraphs is more likely. Similarly, increasing the single qudit error rate results in a larger probability of the error channel not being exactly equivalent to the Pauli-twirled version. As shown in Fig.~\ref{fig:count_sym}, for a single-qudit error rate $p=0.04$, up to $n_h\approx n_v\approx 70$, the probability of not being equivalent to the Pauli-twirled error model is bounded by $p_{\mathrm{not-Pauli-twirled}}\leq 0.04$. For a larger single qudit error rate $p=0.05$, if we require $p_{\mathrm{not-Pauli-twirled}}\leq 0.05$, the lattice size has to satisfy $n_h+n_v<100$.

In Fig.~\ref{fig:count_ana}, we demonstrate how the aspect ratio of the surface code, defined as the ratio between the horizontal size $n_h$ and the vertical size $n_v$, changes the probability $p_{\mathrm{not-Pauli-twirled}}$. For fixed $p$ and $n_h+n_v$, a more symmetric lattice corresponds to a larger probability of not being exactly equivalent to the Pauli-twirled error model. This is because the total number of physical qudits in the surface code is $2n_hn_v-n_h-n_v+1$. If $n_h+n_v$ is fixed, $n_h/n_v$ closer to $1$ indicates a larger product of $n_hn_v$. This means more errors for a fixed physical error rate $p$, therefore the error edges are more likely to form a loop. As seen in the red data points in Fig.~\ref{fig:count_ana} (b), for the most asymmetric lattices considered, namely, $(n_h,n_v)=(16,97)$, $(23,90)$, $(94,19)$, $(99,14)$, $p_{\mathrm{not-Pauli-twirled}}<0.05$ is satisfied, even though for more symmetric lattices $p_{\mathrm{not-Pauli-twirled}}>0.05$.

\begin{figure}[t]
\centering
\includegraphics[width=0.5\textwidth]{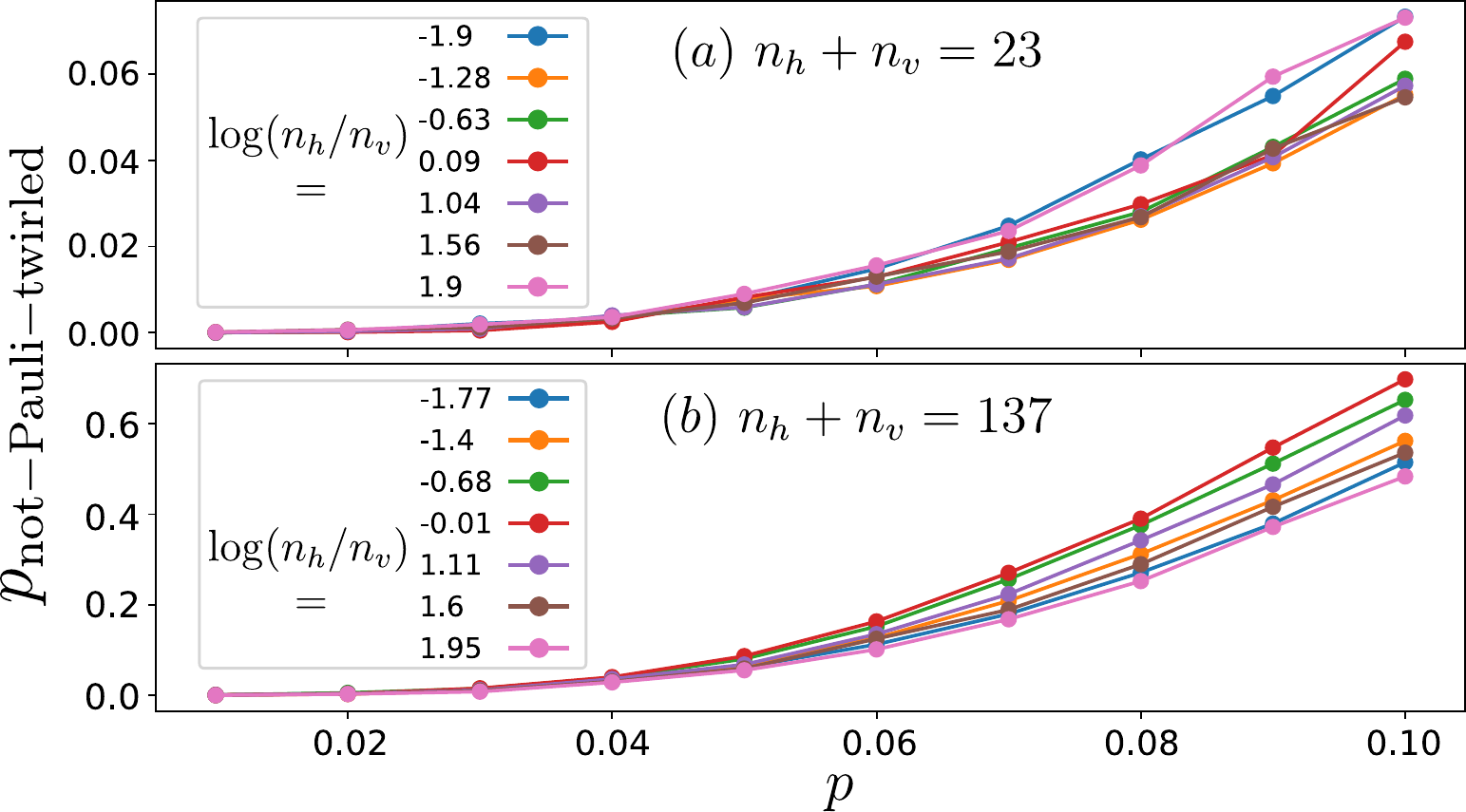}
\caption{Monte-Carlo simulation results of the probability of not being equivalent to the Pauli-twirled error model as a function of the single qudit physical error rate. We fix the aspect ratios of the lattice to different values, each one corresponding to a different color. We consider two fixed values of the perimeter of the lattice, (a) $n_h+n_v=23$, (b) $n_h+n_v=137$. Error bars are small, thus not shown for clarity.}\label{fig:count_aspect}
\end{figure}

In Fig.~\ref{fig:count_aspect}, we show how the probability $p_{\mathrm{not-Pauli-twirled}}$ depends on the single qudit physical error rate $p$. For each pair of $(n_h,n_v)$, which corresponds to a certain value of $n_h+n_v$ and $n_h/n_v$, $p_{\mathrm{not-Pauli-twirled}}$ increases with $p$. This can be expected from the analytical expression Eq.~\eqref{eq:count_ana}, which implies that the dependence contains both $p^3$ scaling coming from the boundaries and $p^4$ scaling coming from the inside of the lattices. In fact, for the cases where $p_{\mathrm{not-Pauli-twirled}}$ is much smaller than the saturation value $1$, the size of the lattice is relatively small such that the boundary contributions are comparable to the non-boundary contributions, or even dominate over the latter ones. In Fig.~\ref{fig:count_aspect} (b), the more symmetric lattices result in larger $p_{\mathrm{not-Pauli-twirled}}$, as discussed before, but in Fig.~\ref{fig:count_aspect} (a), the most asymmetric lattices turn out to lead to larger values of $p_{\mathrm{not-Pauli-twirled}}$. This might be a result of the small size of the lattice, such that logical errors occur at a comparable probability to stabilizers.

\begin{figure}[t]
\centering
\includegraphics[width=0.48\textwidth]{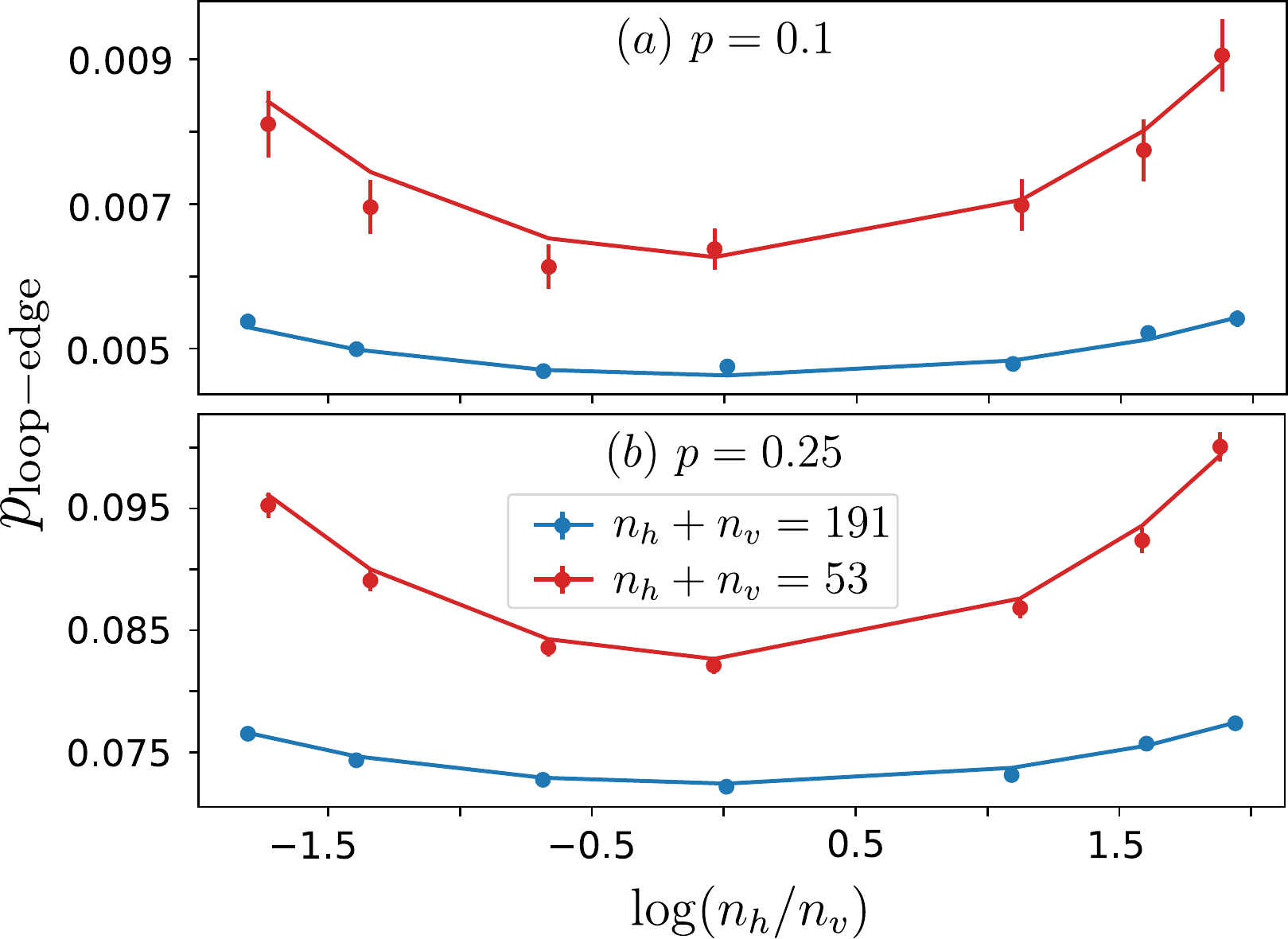}
\caption{The proportion of erroneous qudits that are in loops as a function of the aspect ratio of the lattice, fixing $n_h+n_v=53$ (red) or $n_h+n_v=191$ (blue). Markers are from the Monte-Carlo simulations, and  error bars represent the standard deviation. (a) $p=0.1$, solid lines follow the analytical expression  Eq. (3) in the main text. (b) $p=0.25$, solid lines follow  Eq. (3) in the main text with $4p^3$ replaced by the correction term $4.745p^{3.057}$.}\label{fig:Ratio_ana}
\end{figure}

The value of $p_{\mathrm{not-Pauli-twirled}}$ is useful in setting the upper bound of the logical error rate. To be specific, suppose we have taken $N$ samples, of which $(1-p_{\mathrm{not-Pauli-twirled}})N$ samples have forest error subgraphs and $p_{\mathrm{not-Pauli-twirled}}N$ samples have non-forest error subgraphs. Suppose the logical error rate for forest graphs, $p_l$, can be straightforwardly simulated. Then the largest possible number of samples that cannot be error-corrected is $(1-p_{\mathrm{not-Pauli-twirled}})p_lN+p_{\mathrm{not-Pauli-twirled}}N$, where we have assumed that all the samples with non-forest error subgraphs cannot be error-corrected. The logical error rate without the Pauli twirling approximation is thus upper bounded by  $p_l-p_{\mathrm{not-Pauli-twirled}}p_l+p_{\mathrm{not-Pauli-twirled}}$.

\section{Dependence of the proportion of erroneous qudits that are in loops on the lattice asymmetry}

In the main text, we have shown how $p_{\mathrm{loop-edge}}$ changes with $n_h$ for the symmetric lattices (see main text Fig. 2). Here we demonstrate the dependence of $p_{\mathrm{loop-edge}}$ on the aspect ratio $n_h/n_v$. This is important for small lattices where the contributions from the boundaries are comparable to the contributions from the bulk. Examples are plotted in Fig.~\ref{fig:Ratio_ana}. For a small $p=0.1$, the analytical formula in Eq. (3) in the main text agrees well with the numerical simulations. For a larger $p=0.25$, including the correction as shown in Fig. 3 in the main text leads to a better fit to the simulation data. As the product $n_hn_v$ is in the denominator in Eq. (3) in the main text, a more symmetric lattice results in a smaller value of $p_{\mathrm{loop-edge}}$.

\section{maximal one-dimensional span of a loop for asymmetric lattices}

\begin{figure}[t]
\centering
\includegraphics[width=\textwidth]{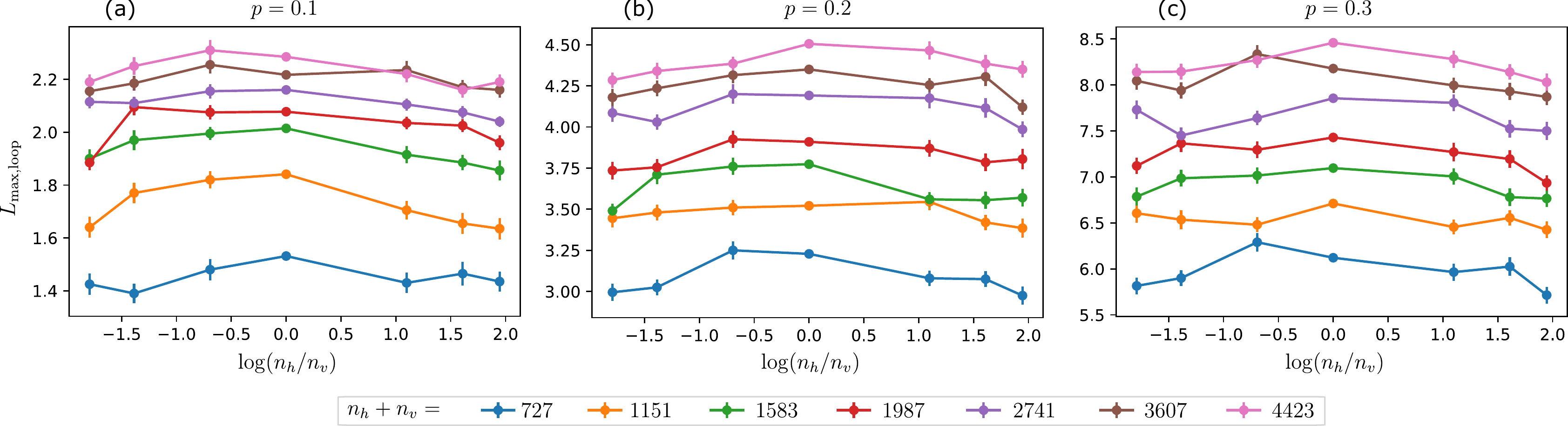}
\caption{The maximal one-dimensional span of a loop as a function of the lattice aspect ratio $n_h/n_v$, fixing the single qudit physical error rate $p$ and the perimeter $n_h+n_v$. Data points are from Monte-Carlo simulations,
where each data point consists of 200 samples for the asymmetric cases and 2000 samples for the symmetric cases. Error bars
represent the standard deviation.}\label{fig:maxLoopAsym}
\end{figure}

In the main text Fig. 4, we have demonstrated how the maximal one-dimensional span of a loop for a symmetric lattice increases with the lattice size. Here we show how it depends on the asymmetry of the lattice in Fig.~\ref{fig:maxLoopAsym}. For the considered large values of $n_h+n_v$, we find that increasing $p$ leads to a larger $L_{\mathrm{max,loop}}$, and increasing $n_h+n_v$ leads to a larger $L_{\mathrm{max,loop}}$ as well, but $L_{\mathrm{max,loop}}$ is relatively insensitive to the value of $\log(n_h/n_v)$ if both $p$ and $n_h+n_v$ are fixed. This is due to the slow logarithmic growth of $L_{\mathrm{max,loop}}$ with respect to the lattice size as illustrated in the main text Fig. 4, effectively suppressing the difference in $n_hn_v$ for fixed $n_h+n_v$ and changing $n_h/n_v$.

\begin{figure}[t]
\centering
\includegraphics[width=0.5\textwidth]{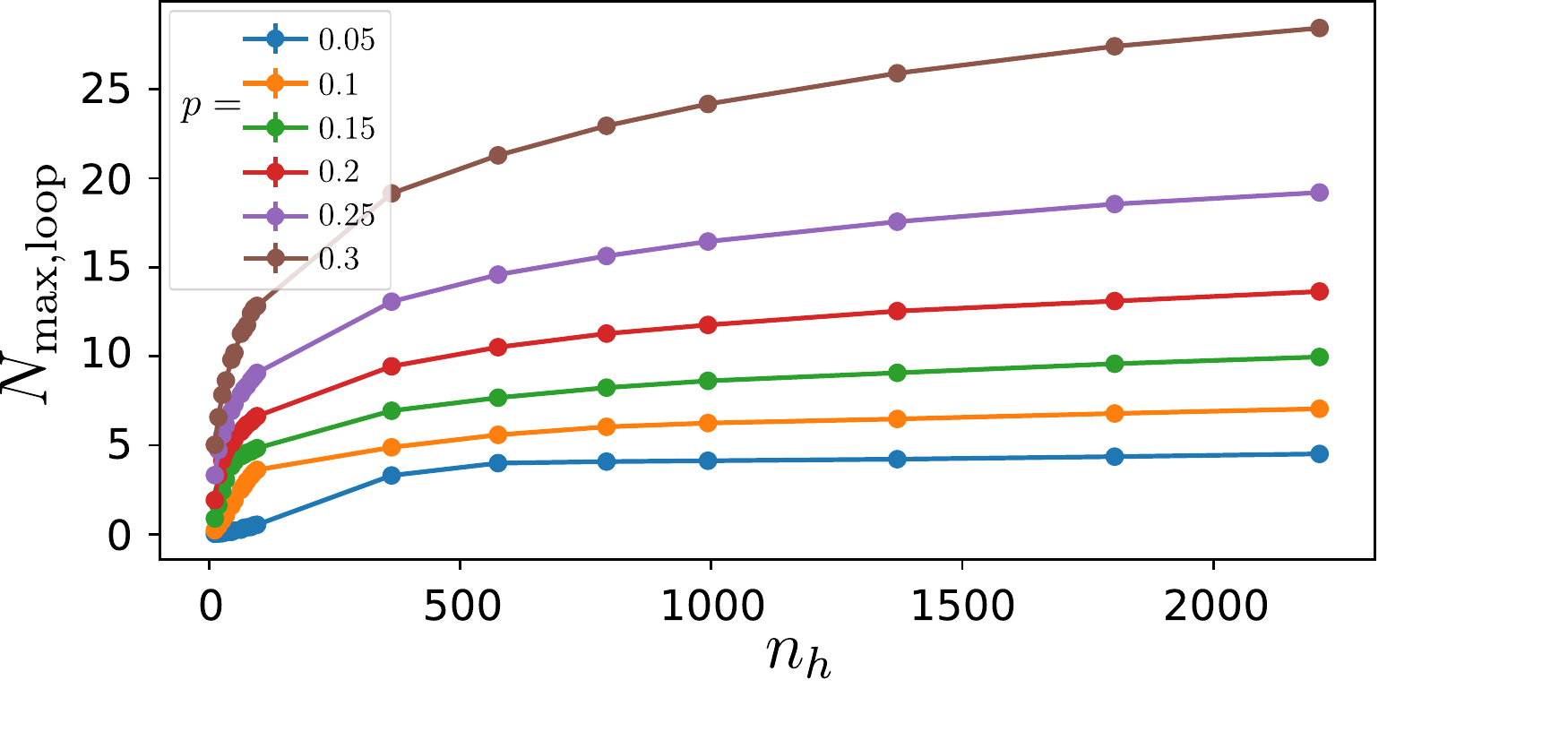}
\caption{The maximal total number of nodes in a loop as a function of the lattice size $n_h$ for almost symmetric lattices $n_h = n_v \pm 1$, obtained from the Monte-Carlo simulations,
where each data point consists of 2000 samples and error bars
represent the standard deviation. Different colors correspond
to different single qudit physical error rates.}\label{fig:maxLoopNum}
\end{figure}

\section{maximal total number of nodes in a loop}

Similar to the one-dimensional span of a loop in the main text, we can define another quantity to describe the size of a loop, namely, the total number of nodes in a loop, $N_{\mathrm{max,loop}}$. The results for symmetric lattices are shown in Fig.~\ref{fig:maxLoopNum}. The slow increase of $N_{\mathrm{max,loop}}$ as a function of $n_h$ follows the same trend as in Fig. 4(a) in the main text. This is because for symmetric lattices the loops are expected to be symmetric in the horizontal and vertical directions, resulting in a fixed relation between  $N_{\mathrm{max,loop}}$ and $L_{\mathrm{max,loop}}$.